\documentclass[prb, twocolumn, showpacs]{revtex4}
\usepackage{graphicx}
\newcommand{\rmi}{\mathrm{i}}
\newcommand{\sgn}{\mathrm{sign}}
\begin{document}
\title{Cooperative phenomenon in a rippled graphene: Chiral spin guide}
\author{M. Pudlak,$^1$ K.N. Pichugin,$^2$
and R.G. Nazmitdinov$^{3,4}$}
\affiliation{$^1$Institute of Experimental
Physics, 04001 Kosice, Slovak Republic\\
$^2$Kirensky Institute of Physics, 660036 Krasnoyarsk, Russia\\
$^3$Departament de F{\'\i}sica,
 Universitat de les Illes Balears, E-07122 Palma de Mallorca, Spain\\
$^4$Bogoliubov Laboratory of Theoretical Physics,
Joint Institute for Nuclear Research, 141980 Dubna, Russia}

\begin{abstract}
We analyze spin scattering in ballistic transport of electrons through
a ripple at a normal incidence of an electron flow. The model of a ripple
consists of a curved graphene surface in the form of an arc of a circle connected
from the left-hand and  right-hand sides to two flat graphene sheets.
At certain conditions the curvature induced spin-orbit coupling
creates a transparent window for incoming electrons  with one spin polarization
simultaneously with a backscattering of those with opposite polarization.
This window is equally likely transparent for electrons with spin up and spin
down that move in opposite directions.
The spin filtering effect being small in one ripple
becomes prominent with the increase of $N$
consequently connected ripples   that
create a graphene sheet of the sinusoidal type. We present
the analytical expressions
for spin up (down) transmission probabilities as a function of $N$ connected
ripples.
\end{abstract}
\date{\today}
\maketitle

\section{Introduction}
The extraordinary properties of graphene have attracted enormous experimental and
theoretical attention for a decade (see e.g. Refs.\onlinecite{1,2}).
Graphene being a zero-gap semiconductor has a band structure
described by a linear dispersion relation at low energy, similar
to massless Dirac-Weyl fermions. Such a band structure leads to exceptionally
high mobility of charged carriers.
A question of possible mechanisms that would allow us to throttle the mobility
and, consequently, to control a conductivity is
a topical subject in graphene physics,
due to its fundamental as well as technological significance.

Among various mechanisms that might affect the mobility, the scattering that could be
induced by a ripple (see, for example, discussion in Ref.\onlinecite{kg}) appears
to be the most natural one, since graphene sheets
are not perfectly flat. Moreover, periodic ripples can be created and controlled
in suspended graphene, in particular, by thermal treatment \cite{lau} and
by placing graphene in a especially prepared substrate.
Indeed, curvature of the surface affects the $ \pi $ orbitals that
determine the electronic properties of graphene. It results in
enhancement of spin-orbit coupling that could serve as a source of spin
scattering.
We recall  that the intrinsic (intraatomic) spin-orbit interaction in  flat graphene
is weak \cite{min,1,2}.
It makes  spin decoherence in such a material 
weak as well, i.e., scattering due to disorder is supposed to be unimportant.
In order to get deep insight into the nature of  curvature induced scattering,
it is desirable to elucidate among many questions the basic one: What are
the distinctive features of  curvature induced spin-orbit coupling ?
One can further ask
how to employ these features to
guide an electron transport in a graphene-based system at the theoretical, and, 
quite likely, practical levels.

A consistent approach to introduce the curvature induced spin-orbit coupling (SOC)
in the low energy physics of graphene have been developed by Ando \cite{Ando} and by others
\cite{Ent,Mart,Brat} in the framework of effective mass and tight-binding
approximations.
Recent measurements in ultra clean carbon nanotubes (CNTs) \cite{Ku}, i.e.,
in an extreme form of  curved graphene,
revealed the energy splitting that
can be associated with  spin-orbit coupling.
The measured shifts are compatible with theoretical
predictions \cite{Ando,Brat},  while some features
regarding the contribution of different spin-orbit terms in metallic and
non-metallic CNTs are still debatable
(see, for example, discussion in \cite{Loss,Chico,Jeo,izum,val,kl}).
Nowadays, nevertheless, there is a consensus that for armchair CNTs
one obtains two SOC terms: one preserves the spin symmetry
(a spin projection on the CNT symmetry axis),
while the second one breaks this symmetry
\cite{Ando,izum,val,at1}. Thus, we have a reliable answer to the first question, at least,
for armchair CNTs.
In some previous studies \cite{Ando,Brat,izum,val}
the role played by the second term
was underestimated. In this paper we will attempt to answer
how  full curvature induced SOC, including the second term, could be used to create
a polarized spin current with a high efficiency in a rippled graphene system.

The structure of this paper is as follows. In Sec. II we briefly discuss the explicit expressions
for the eigen spectrum and eigenfunctions of an armchair nanotube with a full
curvature induced  spin-orbit coupling.
By means of these results we introduce a scattering model for one ripple
and extend this model for $N$ continuously connected ripples.
In Sec. III we provide a discussion of our results in terms
of simple estimates.
The main conclusions are summarized in Sec. IV.

\section{Scattering problem}

In order to model a scattering problem on a ripple
 we consider a curved surface in the form of an arc of a circle connected
from the left-hand and right-hand sides to two flat graphene sheets.
The solution for  flat graphene is well
known \cite{1,2}. The solution for a curved graphene surface can be
expressed in terms of the results obtained for  armchair CNTs
in an effective mass approximation, when only the interaction
between nearest neighbor atoms is taken into account \cite{at1}.

\subsection{Low energy spectrum of the armchair nanotube}

Let us recapitulate the major results \cite{at1} in the vicinity of the Fermi level
$E=0$ for a point $K$ in the
presence of the curvature induced spin-orbit interaction in an armchair CNT.
The $y$ axis  is chosen as the symmetry and the quantizations axis.
 In this case the eigenvalue problem is defined as
\begin{equation}
\label{H}
\hat{H}\Psi=
\left(\begin{array}{cc}0&\hat{f}\\
\hat{f}^{\dag}&0
\end{array}
\right)\left(
\begin{array}{c}
F^{K}_{A}\\F^{K}_{B}
\end{array}
\right)=E\left(
\begin{array}{c}F^{K}_{A}\\F^{K}_{B}\end{array}\right)\,,
\end{equation}
with the following definitions:
\begin{eqnarray}
\label{Ha}
& \hat{f}=\gamma
(\hat{k}_{x}-\rmi\hat{k}_{y})+\rmi\frac{\delta\gamma'}{4R}\hat{\sigma}_x(\vec{r})-
\frac{2\delta \gamma p}{R}\hat{\sigma}_{y}\,,\\
& \hat{k}_{x}=-\rmi\frac{\partial}{R\partial\theta}\,,
\hat{k}_{y}=-\rmi\frac{\partial}{\partial y}\,,\nonumber\\
&\hat{\sigma}_x(\vec{r})=\hat{\sigma}_{x}\cos\theta -\hat{\sigma}_{z}\sin\theta\,.
\nonumber
\end{eqnarray}
Here, $\hat{\sigma}_{x,y,z}$ are standard Pauli matrices,
and the spinors of two sub-lattices are
\begin{equation}
F^{K}_{A}=\left(\begin{array}{c}F^{K}_{A,\uparrow}\\F^{K}_{A,\downarrow}\end{array}\right)\,,\quad
F^{K}_{B}=\left(\begin{array}{c}F^{K}_{B,\uparrow}\\F^{K}_{B,\downarrow}\end{array}\right)\,.
\end{equation}
The following notations are used: $\gamma =-\sqrt{3}V_{pp}^{\pi}a/2=\gamma_{0}a$,
$\gamma'=\sqrt{3}(V_{pp}^{\sigma}-V_{pp}^{\pi})a/2=\gamma_{1}a$,  $p=1-3\gamma'/8\gamma$
 (see e.g. Ref.\onlinecite{Ando}).
The quantities $V_{pp}^{\sigma}$ and $V_{pp}^{\pi}$ are the transfer
integrals for $\sigma$ and $\pi$ orbitals, respectively in a flat graphene;
$a=\sqrt{3}d\simeq 2.46$ \AA{}
is the length of the primitive translation vector,
where $d$ is the distance between atoms in the unit cell.

The intrinsic source of the SOC
$\delta = {\Delta}/(3\epsilon_{\pi\sigma})$ is defined as
\begin{equation}
\Delta = \rmi\frac{3\hbar}{4m_e^{2}c^{2}}\langle x_l|\frac{\partial
V}{\partial x} \hat{p}_{y}-\frac{\partial V}{\partial y} \hat{p}_{x}|y_l\rangle\,,
\end{equation}
where $V$ is the atomic potential
and $\epsilon_{\pi\sigma}=\epsilon_{2p}^{\pi}-\epsilon_{2p}^\sigma$.
The energy $\epsilon_{2p}^{\sigma}$ is the energy of
$\sigma$-orbitals, localized between carbon atoms. The
energy $\epsilon_{2p}^{\pi}$ is the energy of $\pi$-orbitals,
directed perpendicular to the curved surface.

By means of the unitary transformation
\begin{equation}
\label{U}
\hat{U}(\theta)=\exp({\rmi\frac{\theta}{2}}\hat{\sigma}_{y})\otimes I\,,
\end{equation}
where $I$ is $2\times2$ unity matrix,
one removes the $\theta$ dependence in Hamiltonian (\ref{H}), transformed in the intrinsic frame,
 and obtains
\begin{eqnarray}
\label{H2}
&& \hat{H}' = \hat{U}(\theta)\hat{H}\hat{U}^{-1}(\theta)=\hat{H}_{\it kin}+\hat{H}_{SOC}\,,\\
&& \hat{H}_{\it kin} = -\rmi\gamma \left(\hat{\tau}_y\otimes I\partial_y+
\hat{\tau}_x\otimes I
\frac{1}{R} \partial_\theta \right)\,,\nonumber\\
&& \hat{H}_{SOC} = -\lambda_y\hat{\tau}_y\otimes\hat{\sigma}_x -
\lambda_x\hat{\tau}_x\otimes\hat{\sigma}_y \,.\nonumber
\end{eqnarray}
Here the operators $\hat{\tau}_{x,y,z}$ are the Pauli matrices
that act on the wave functions of A- and B-sub-lattices (a pseudo-spin space), and
\begin{equation}
\label{lxy}
\lambda_x=\gamma\left(1+4\delta p\right)/(2R)\,,\quad
\lambda_y=\delta\gamma^{\prime}/(4R)
\end{equation}
are the strengths of the SOC terms.
In the Hamiltonian (\ref{H2})
the term  $(\sim \lambda_x)$ conserves, while
the other one $(\sim \lambda_y)$ breaks the spin symmetry in the armchair CNT.

The operator $\hat{J}_{y}$, being an integral of motion $\left[\hat{H},\hat{J}_y\right]=0$,
is defined  in the laboratory frame as
\begin{equation}
\hat{J}_y=I\otimes\left(\hat{L}_y+\frac{\hat{\sigma}_y}{2}\right)=
I\otimes\left(-\rmi\partial_\theta+\frac{\hat{\sigma}_y}{2}\right)\,,
\end{equation}
while in the intrinsic frame it is
\begin{equation}
\hat{J}_y\rightarrow
\hat{J}_y'=\hat{U}\hat{J}_y\hat{U}^{-1}=
I\otimes\left(-\rmi\partial_\theta\right).
\end{equation}
This integral allows to present the  wave functions  as
\begin{equation}
\label{F}
 F'(\theta,y)=e^{\rmi m\theta}e^{\rmi k_y y}\Psi
=e^{\rmi m\theta}e^{\rmi k_y y}\left(\begin{array}{c}A\\B\\C\\D\end{array}\right)\,.
\end{equation}
These wavefunctions are also the eigenfunctions of the other integral of motion, the operator
$\hat{k}_y'\equiv\hat{k}_y$. Here,
$m=\pm 1/2,\pm 3/2,....$ is an eigenvalue of
the angular momentum operator
$\hat{J}_y'$.
For the components of the eigenvector $ F'(\theta,y)$
the relations $|A|=|D|$ and $|B|=|C|$ are fulfilled
at real values of $m$ and $k_y$.

Solving the eigenvalue problem $\hat{H}'F'=E F'$, one obtains the eigen spectrum
\begin{eqnarray}
\label{ener}
&E=\pm E_{m,q}\,,\quad
E_{m,q}= \sqrt{t_m^2+t_y^2+\lambda_y^2+\lambda_x^2+2 D_{m,q}}\,,\nonumber\\
&D_{m,q} = q \sqrt{\lambda_x^2\left(t_m^2+\lambda_y^2\right)+t_y^2\lambda_y^2}\,,\quad q=\pm 1\,,
\end{eqnarray}
where $t_m=\gamma m/R$, $t_y=\gamma k_y$.

\subsection{Scattering model for one ripple}

\begin{figure}
\begin{center}
\includegraphics[width=8cm]{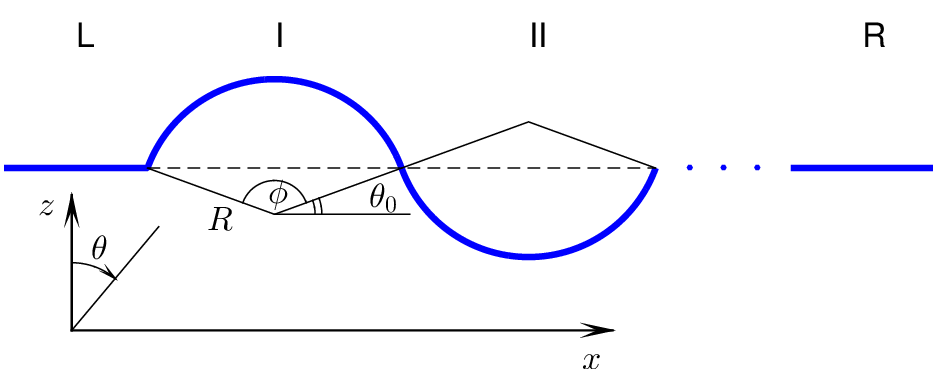}
\caption{The rippled graphene system.}
\label{fig1}
\end{center}
\end{figure}

Keeping in mind a discussion that will be given hereafter, we analyze the following
geometry (see the construction profile on Fig.\ref{fig1}).
It consists of one arc of a circle that is connected
from the left-hand side to a flat graphene sheet. This (direct) arc is continuously connected to
the inverse arc of the same radius that is connected to the right-hand flat graphene sheet.
We put the origin of the coordinate at the center of the direct arc of the circle.

To give  better insight into the scattering phenomenon in our model of a ripple,
we study first only the direct arc of the circle connected to two
flat surfaces.
Two flat surfaces are:
i)the region L, defined in the intervals
$-\infty <x< -R\cos\theta_{0}$;
the region R, defined in the intervals
$R\cos\theta_{0}< x < \infty$.
The region I is a part of a
nanotube of radius $R$, defined as $-R\cos\theta_{0} <x< R\cos\theta_{0}$.
At $\theta_{0}=0$, the
ripple is a half of the nanotube, while at
$\theta_{0}= \pi /2$ the ripple does not exist.
For the sake of analysis we  introduce the angle $\phi=\pi-2\theta_0$.
To describe the scattering phenomenon one has to define
wave functions in different regions: flat  (L,R) and
curved (I) graphene surfaces.

Regions L and R are described by the Hamiltonian
\begin{equation}
\label{HF}
\hat{H}_0=
\gamma \left(\hat{\tau}_x k_x+\hat{\tau}_y k_y\right)\otimes I
\end{equation}
that does not mix  spin components.
For the sake of simplicity, we consider
the electron motion at the normal incidence,
with the electron wave vector $\vec{k}=(k_x,0)$.
One solves the stationary Schr\"{o}dinger equation
$\hat{H}_0\Psi=E_0\Psi$
and obtains
the corresponding eigenstates
\begin{eqnarray}
&E_0=\pm \gamma |k_x|\,,\\
&\Psi=\exp{(\rmi k_xx)}\Psi_0^{\sigma}(k_x)\,,\\
&\label{eigI}
\Psi_0^{\sigma}(k_x)=
\frac{1}{2}\left(
\begin{array}{c}
\sgn(\gamma k_x E_0)\Phi_0^{\sigma}\\
\Phi_0^{\sigma}
\end{array}\right),\\
&\label{eigI1}
\hat{\sigma}_y\Phi_0^{\sigma}=
\sigma \Phi_0^{\sigma}\,,
\Phi_0^{\sigma}=\left(\begin{array}{c}1\\ \rmi \sigma\end{array}\right),
\sigma=\pm 1.
\end{eqnarray}

Evidently, the wave functions in regions L, R, can be written as
a superposition of all possible solutions for  flat graphene.
To proceed, with the aid of eigenspinors (\ref{eigI}),  (\ref{eigI1}), we introduce an auxiliary
matrix $\hat{M}_0$ $(4\times 4)$ for a given value of energy at the normal incidence
\begin{equation}
\label{Mf}
\hat{M}_0=\left(
\Psi_0^{+1}(K_x)\,, \Psi_0^{-1}(K_x)\,,
\Psi_0^{+1}(-K_x)\,, \Psi_0^{-1}(-K_x)
\right)\,.
\end{equation}
Here, we define the variable $K_x=\sgn(\gamma E_0)|k_x|$
to ensure that  the first two columns of the matrix $M_0$
correspond to eigenstates that move in a positive $x$-direction, while  the last two columns
correspond to eigenstates that move in a negative $x$-direction.

The matrix $\hat{M}_0$ is
unitary, i.e., $\hat{M}_0^{-1}=\hat{M}_0^\dag$.
It allows us to define a general form of
the wave function $\Psi_{L,R}$
for a  flat graphene
\begin{equation}
\Psi_{L,R}(x)=\hat{M}_0\exp\left(\rmi \hat{K} (x-x_{L,R})\right) C_{L,R}\,,
\end{equation}
where
$\hat{K}={\rm diag}(K_x,K_x,-K_x,-K_x)$ is a diagonal matrix,
$x_{L,R}$ are $x$-coordinates where  flat and curved  surfaces are
connected, and
$C_{L,R}$ are corresponding vectors with
four unknown yet, normalized coefficients in each region.
Note that we do not consider inelastic scattering. Therefore,
since the electron energy is conserved, we use the same vector
$\vec{k}=(k_x,0)$  for the left and right flat graphene surfaces.

\begin{figure}
\includegraphics[height=6cm,clip=]{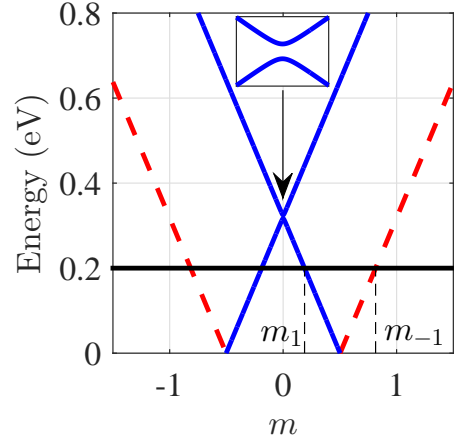}
\caption{(Color online) The spectrum (\ref{mval}) at
$k_y=0$ as a function of the quantum number $m$.
Dashed (red) and  solid (blue) lines are associated with states characterized by the
quantum number $m_{s=-1}$ and $m_{s=+1}$, respectively.
The values of
$\pm m_{s}$ at the energy $E_F=0.2$eV (solid horizontal line) are indicated by
vertical dashed lines.
The parameters of calculations are $R=10$\AA, $\delta=0.01$, $p=0.1$,
$\gamma=\frac{9}{2}1.42$\AA eV, $\gamma'=\gamma \frac{8}{3}$,
$\lambda_x=\frac{\gamma}{R}(1/2+2\delta p)=0.32$eV,
$\lambda_y=\frac{\delta\gamma'}{4R}=0.0043$eV.
The arrow indicates the gap displayed on the insert.}
\label{fig_spectrum}
\end{figure}

For the curved surface we use eigenspinors of the Hamiltonian (\ref{H2}).
The general form of these eigenspinors is defined
in the intrinsic frame \cite{at1}.  Therefore, we apply
the inverse transformation (\ref{U}) to these eigenspinors  in order
to analyze the scattering problem in the laboratory frame.
At $k_y=0$ the spectrum (\ref{ener}) and the eigenspinors
 are particularly simple
\begin{eqnarray}
\label{mval}
&E_A=\pm\sqrt{t_m^2+\lambda_y^2}+s\lambda_x\,,\quad s=\pm 1\,,\\
&\Psi=\exp{(\rmi m\theta)}\;\Psi_A^{s}(t_m)\,,\\
\label{eigII}
&\Psi_A^{s}(t_m)=\left(
\begin{array}{c}
-s\Phi_A^{s}(t_m)\\
\sigma_y\Phi_A^{s}(t_m)
\end{array}\right)\,,\\
&\label{eig3}
\Phi_A^{s}(t_m)=\exp\left({-\rmi\frac{\theta}{2}}\hat{\sigma}_{y}\right)
\left(
\begin{array}{c}
-\rmi t_m\\
\lambda_y-(s E_A-\lambda_x)
\end{array}\right)\,.
\end{eqnarray}
Note that energies in  flat graphene
and in a curved surface are different $E_0=E_A+A(\frac{a}{R})^2$
(see details in Ref.\onlinecite{Pudlak}).
This effect is caused
by different hybridizations of $\pi$ electrons in  flat
graphene and a graphene--based system with 
curvature.
In particular,  $A=5/6, 7/12$ (eV) in the armchair and zig-zag nanotubes, respectively.

At a fixed energy of the electron flow $E_0\Leftrightarrow E_A$, Eq.({\ref{mval}) yields four possible values of
the quantum number $m$
\begin{equation}
\label{mv}
m\Rightarrow m_{s}=\pm\frac{R}{\gamma}\sqrt{(s E_A-\lambda_x)^2-\lambda_y^2},\quad s=\pm 1.
\end{equation}
Since the angular momentum is not longer the integral of motion, we have to
consider the mixture of eigenfunctions with all possible values of ${m}$  at a given energy.

As an example of the spectrum  (\ref{mval}), a few positive energy branches as a function
of the quantum number $m$ are shown in Fig.2.
The branches are distinguished by the index $s=\pm 1$.
There is an anti-crossing effect between  energy states characterized
by the same $m_{s=+1}$ quantum number.
This anticrossing is brought about by the interaction ($\sim \lambda_y$)
that breaks the spin symmetry   (see Sect.IIA)
in the curved graphene surface.
It results in a gap of $=2\lambda_y$
near $E_A=\lambda_x$ indicated by the
arrow (see the inset on Fig.\ref{fig_spectrum}).
Similar gap occurs near
$E_A=-\lambda_x$ for the $m$-states with index $s=-1$.
As a consequence of these
gaps,  evanescent modes arise at energies
$\lambda_x-\lambda_y<|E|<\lambda_x+\lambda_y$ in our system.
For the sake of illustration the positive spectrum (\ref{mval}) of $m$-states is crossed by the
horizontal line that mimics the Fermi energy. The crossing points
determine quantum
numbers $m$ that have non-quantized values when the curved surface
(arc of circle) is connected to the flat one.

With the aid of eigenspinors (\ref{eigII}), (\ref{eig3}), and the unitary
transformation (\ref{U}), we introduce an auxiliary
matrix for a given value of energy at a curved surface
\begin{eqnarray}
\label{Mr}
&\hat{M}_A(\theta)=\nonumber\\
&\left(
\Psi_A^{1}(t_{m_1})\,, \Psi_A^{-1}(t_{m_{-1}})\,,
\Psi_A^{1}(-t_{m_1})\,, \Psi_A^{-1}(-t_{m_{-1}})
\right)=\nonumber\\
&=U(-\theta)\hat{M}_A(0).
\end{eqnarray}
As a result, in  region I  the wave function  can be written as
a superposition of all solutions for a curved surface in the form
$\Psi_I(\theta)=\hat{M}_A(\theta)\exp(\rmi \hat{m}\theta) C_{I}$.
Here, $C_{I}$ is a vector of four unknown coefficients,
$\hat{m}={\rm diag}(m_{1},m_{-1},-m_{1},-m_{-1})$ is a
diagonal matrix.

The overlap of eigenspinors of the flat and bended regions
can be readily calculated with
the aid of Eqs.(\ref{eigI}), (\ref{eigII}), which results in
\begin{eqnarray}
\label{overlap}
&\left(\Psi_0^{\sigma}\right)^\dag \left(\Psi_A^{s}\right)\simeq
\left(-\sgn(\gamma k_x E_0) s+\sigma\right)\times\\
&\times\left(\Phi_0^{\sigma}\right)^\dag
\exp\left({-\rmi\frac{\theta}{2}}\hat{\sigma}_{y}\right)
\left(\Phi_A^{s}\right)\nonumber.
\end{eqnarray}
Evidently, the overlap is zero at $\sigma=\sgn(\gamma k_x E_0) s$.
Note that already this result implies that some of the four channels between the flat and
curved regions could be closed.

Matching  the wave functions at the boundaries of regions L, I, and R,
for an incoming electron flow from the left-hand side,
leads us to the following equations
\begin{eqnarray}
\label{match}
&&\Psi_L(x_L)=\Psi_I(-\phi/2)\Rightarrow\nonumber\\
&&\hat{M}_0\left(\begin{array}{c}\Phi_{in}\\r\end{array}\right)=
\hat{M}_A(-\phi/2) C_{I}\,,\\
\label{match1}
&&\Psi_R(x_R)=\Psi_I(+\phi/2)\Rightarrow\nonumber\\
&&\hat{M}_0\left(\begin{array}{c}t\\0\end{array}\right)=
\hat{M}_A(+\phi/2) \exp(\rmi \hat{m}\phi) C_{I}\,.
\end{eqnarray}
We recall that the angles $\theta_0$ and $\phi$ determine the $x_{L,R}$ coordinates:
$x_L=R\cos (\theta_0+\phi)$, $x_R=R\cos \theta_0$. Here, $t=\left(\begin{array}{c}t(L)_{\uparrow}^{in}\\ t(L)_{\downarrow}^{in}\end{array}\right)$
and $r=\left(\begin{array}{c}r(L)_{\uparrow}^{in}\\ r(L)_{\downarrow}^{in}\end{array}\right)$
are transmission and reflection coefficients, respectively, for incoming electron either
with a spin up $|\uparrow\rangle\equiv\left(\begin{array}{c}1\\\rmi\end{array}\right)$
 or with a spin down $|\downarrow\rangle \equiv
\left(\begin{array}{c}1\\-\rmi\end{array}\right)$.

Solutions of Eqs.(\ref{match}) (and similar equations
for an incoming electron flow from the right-hand side)
yield the following probabilities
\begin{eqnarray}
\label{S1}
|t(L)_{\uparrow}^{\uparrow}|^2=|t(R)_{\downarrow}^{\downarrow}|^{2}=
\frac{1}{1+(z_{-1})^2}\,\\
\label{S2}
|t(L)_{\downarrow}^{\downarrow}|^2=|t(R)_{\uparrow}^{\uparrow}|^{2}=
\frac{1}{1+(z_{+1})^2}\,\\
\label{S3}
|r(L)_{\downarrow}^{\uparrow}|^2=|r(R)_{\uparrow}^{\downarrow}|^{2}=
1-\frac{1}{1+(z_{-1})^2}\,\\
\label{S4}
|r(L)_{\uparrow}^{\downarrow}|^2=|r(R)_{\downarrow}^{\uparrow}|^{2}=
1-\frac{1}{1+(z_{+1})^2}\,.
\end{eqnarray}
Here, we have also introduced the variable $z_{s}$
\begin{equation}
\label{z}
z_{s}=\frac{\lambda_y\sin(m_{s}\phi)}{t_{m_{s}}}=
\frac{\lambda_yR}{\gamma}\times\frac{\sin(m_{s}\phi)}{m_{s}}
\,,\quad s=\pm 1\,,
\end{equation}
related to the characteristics of the curved surface (see Sec.IIA).

Evidently, there is no  backscattering for incoming electrons, if
$\lambda_y=0$ [see Eqs.(\ref{S3})-(\ref{z})].
However, at $\lambda_y\neq0$ backscattering with a spin inversion takes place.
The reflection probabilities without the spin inversion are
$|r(L)_{\uparrow}^{\uparrow}|^2=|r(L)_{\downarrow}^{\downarrow}|^2=0$.
The same is true for the transmission probabilities with a spin  inversion, i.e.,
$|t(L)_{\downarrow}^{\uparrow}|^2=|t(L)_{\uparrow}^{\downarrow}|^2=0$.
Thus,  backscattering  with a spin inversion is nonzero
in the ripple due to the curvature induced SOC produced by the
$\lambda_y$-term. In addition, incoming
electrons with different spin orientations choose different channels
(different $m_s, s=\pm1$).

The maximum transmission probability $|t(L)_{\uparrow}^{\uparrow}|^2=1$
takes place at the condition
\begin{equation}
\label{eff1}
m_{-1}\phi_c=\pi n\,, n=1,2,\dots\,,
\end{equation}
[see Eqs.(\ref{S1}),(\ref{z})].
Evidently, this probability becomes dominant
at the minimum transmission $|t(L)_{\downarrow}^{\downarrow}|^2$.
The lowest minimum of the
transmission $|t(L)_{\downarrow}^{\downarrow}|^2$ occurs at the condition $E_A=\lambda_x$,
when $m_{+1}$ becomes imaginary [see Eq.(\ref{mv})].
In other words, the propagating mode transforms to the evanescent mode
for the channel $m_{+1}$.
Taking into account the condition $E_A=\lambda_x$ in Eq.(\ref{mv}),
one obtains the critical angle of the curved surface (in form of the
arc) for  a maximum of spin up filter efficiency
\begin{equation}
\label{theta_c}
\phi_c=\frac{\pi n}{m_{-1}}=
\frac{\pi n \gamma}{R\sqrt{4\lambda_x^2-\lambda_y^2}}\,,
\end{equation}
where the SOC strengths $\lambda_{x,y} $ are defined by Eq.(\ref{lxy}).
For parameters listed in the caption of Fig.\ref{fig_spectrum} we have
$|\phi_c|\approx 0.996\pi$($ n=1$).
For the same critical angle $\phi_c$ and
$E_A=-\lambda_x$ we obtain a maximum for the spin down filter efficiency,
when $m_{-1}$ becomes imaginary.

Thus, there are different channels for the spin up and spin down
electron (hole) flows. Note that the deviation from
the energy value $E_A=\pm\lambda_x$
could produce the equal transmission for spin up and spin down electrons
(see Fig.\ref{fig_RT200}.a).
Therefore, it is important to choose the energy $|E_A|$  to be in
the close vicinity of the energy value $\simeq\lambda_x$.
For the considered parameters the filter efficiency is, however,  small.
So far this result has met with only limited success.

\subsection{Scattering model for $N$ ripples}

To increase the efficiency  we suggest  connecting the bent parts sequentially,
as shown in Fig. \ref{fig1}. In particular, the construction with the
direct+inverse arcs (with the same angle $\phi$) transforms
Eqs.(\ref{match})(\ref{match1}), to the forms
\begin{eqnarray}
&&\hat{M}_0\left(\begin{array}{c}\Phi_{in}\\r\end{array}\right)=
\hat{M}_A(-\phi/2) C_{I}\,,\nonumber\\
&&\hat{M}_0\left(\begin{array}{c}t\\0\end{array}\right)=
\hat{M}_A(\pi-\phi/2) \exp(-\rmi \hat{m}\phi)\times\\
&&\hat{M}_A^{-1}(\phi/2-\pi)\hat{M}_A(+\phi/2) \exp(+\rmi \hat{m}\phi) C_{I}\,.\nonumber\\
\end{eqnarray}

Since  in the inverse arc the phases, accumulated from the point of
connection with the direct arc to the
point of connection with a straight line  ( flat graphene), have
a sign opposite  that of the first one, we use $\exp(\pm\rmi \hat{m}\phi)$.

Matching  the wave functions at the boundaries of regions L, I, II, and R,
for electron  coming from the left-hand (L) and  right-hand (R) sides of the construction,
leads us to the following nonzero probabilities

\begin{eqnarray}
&&|t(L)_{\uparrow}^{\uparrow}|^2=
\left[\frac{1}{1+2(z_{-1})^2}\right]^2=|t(R)_{\downarrow}^{\downarrow}|^{2}\,,\\
&&|t(L)_{\downarrow}^{\downarrow}|^2=
\left[\frac{1}{1+2(z_{+1})^2}\right]^2=|t(R)_{\uparrow}^{\uparrow}|^{2}\,,\\
&&|r(L)_{\downarrow}^{\uparrow}|^2=
1-|t(L)_{\uparrow}^{\uparrow}|^2=|r(R)_{\uparrow}^{\downarrow}|^{2}\,,\\
&&|r(L)_{\uparrow}^{\downarrow}|^2=
1-|t(L)_{\downarrow}^{\downarrow}|^2=|r(R)_{\downarrow}^{\uparrow}|^{2}\,.
\end{eqnarray}

Thus, the transmissions through one and two (direct+inverse) arcs
are accompanied by the inverse backscattering. The considered cases imply that
the larger the number of arcs is, the stronger the inverse backscattering is
for one of the spin components.

Following the recipe described in Ref.\onlinecite{dat} (interfering Feynman paths),
 and combining S-matrices
for $N$ connected arcs, we obtain

\begin{eqnarray}
\label{SN1}
|t(L)_{\uparrow}^{\uparrow}|^2=\left[\frac{2}{{C_{-1}^{(+)}}^N+{C_{-1}^{(-)}}^N}\right]^2
=|t(R)_{\downarrow}^{\downarrow}|^2\,,\\
\label{SN2}
|t(L)_{\downarrow}^{\downarrow}|^2=\left[\frac{2}{{C_{+1}^{(+)}}^N+{C_{+1}^{(-)}}^N}\right]^2
=|t(R)_{\uparrow}^{\uparrow}|^2\,.
\end{eqnarray}
Here, the variable $C_{s}^{(\pm)}$ is defined as
\begin{equation}
C_{s}^{(\pm)}=\sqrt{1+(z_{s})^2}\pm z_{s}\,,
\quad s=\pm 1.
\label{cs}
\end{equation}

Evidently, at $z_{s}=0$ the transmission probability is $1$ for any
number of arcs, while $z_{s}\neq 0$ leads to the decrease of the transmission
probability with the  increase of  the number of arcs.
The suppression is, however, different
for various transmission probabilities
due to their different dependence on
the  quantum number $m_s$.

As shown above,  conditions (\ref{eff1}), (\ref{theta_c}), determine
the dominance, in particular, of the transmission probability of
spin up incoming electrons at $E_A>0$.
Indeed, a set $N\gg1$ of an exact replica of the consistently connected ripples
(see Fig.\ref{fig1}) does not affect this dominance $(=1)$
for the $m_{s=-1}$ channel.  However, this set
suppresses the spin down transmission probability for the $m_{s=+1}$ channel
that is proportional to $x<1\Rightarrow x^N\rightarrow0$.

We would like to point out that Eqs.(\ref{SN1}),(\ref{SN2}), are valid for odd and even
number of consistently connected ripples. In our model the only requirement is 
that the direct ripple has to be connected to the inverse one, the inverse ripple to the direct one etc.
From our consideration it follows that, if at a certain energy, for example, $E_A>0$
there is a high transmission probability for {\it spin-up} electrons from the left side of our
system, one obtains the same magnitude for the transmission probability
for {\it spin-down} electrons from the right side.

\section{Discussion}
\begin{figure}
\includegraphics[height=5cm,clip=]{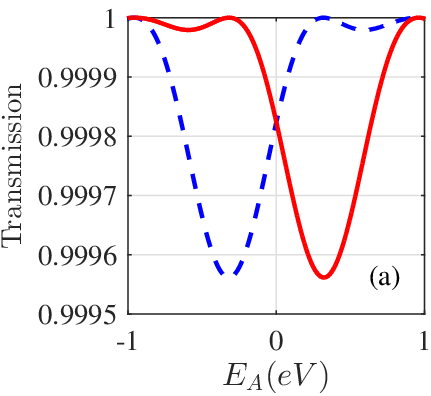}
\includegraphics[height=5cm,clip=]{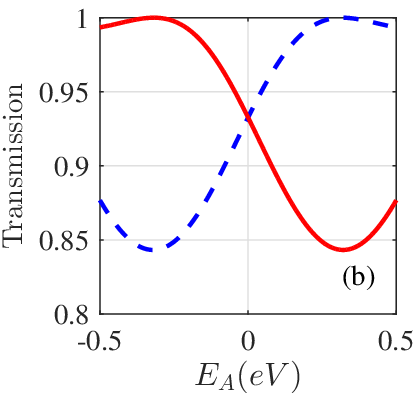}\\
\includegraphics[height=5cm,clip=]{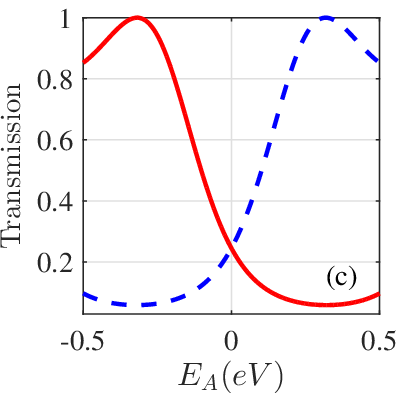}
\includegraphics[height=5cm,clip=]{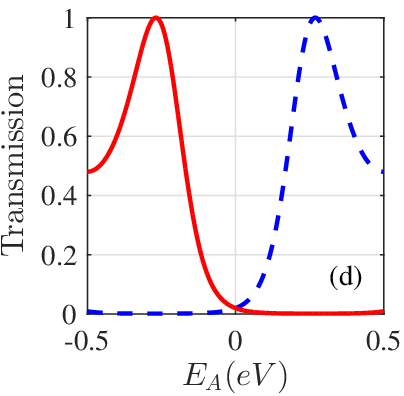}
\caption{(Color online) Dependence of transmission
probabilities $|t(L)_{\uparrow}^{\uparrow}|^2$ (blue, dashed lines) and
$|t(L)_{\downarrow}^{\downarrow}|^2$ (red, solid   lines)
on the energy $E_A$ at $k_y=0$ for
1 (a), 20 (b), 100 (c) and 200 (d) sequentially connected
ripples ($\pi$-arcs).
The parameters are the same as in Fig.\protect\ref{fig_spectrum}.}
\label{fig_RT200}
\end{figure}

\subsection{N-factor}

To obtain a simple picture of the physics behind the enhancement of
the spin filtering effect, let us consider the transmission at
the energy $E_A\simeq\lambda_x$,
when $m_{+1}$ becomes imaginary [see Eq.(\ref{mv})] and
the propagating mode transforms to the evanescent mode
for the channel $m_{+1}$.
In light of Eqs.(\ref{mv}),(\ref{lxy}), one obtains
\begin{equation}
\label{mp}
m_{+1}=i\frac{R}{\gamma}\lambda_y=i x\,, x=\frac{\delta \gamma^\prime}{4\gamma}\approx0.007
\end{equation}
As a result, the variable $z_+$ (Eq.(\ref{z})) transforms in the form
\begin{equation}
z_+\simeq x\phi \ll 1\,.
\label{zp}
\end{equation}
Taking into account Eqs.(\ref{cs}-\ref{zp}), one can readily estimate
that at $N\gg1$
\begin{eqnarray}
{C_{+1}^{(+)}}^N+{C_{+1}^{(-)}}^N&\simeq& 2+(Nx\phi)^2\Rightarrow\\
\Rightarrow|t(L)_{\downarrow}^{\downarrow}|^2&\approx&\left[\frac{2}{2+(Nx\phi)^2}\right]^2
\end{eqnarray}
With our choice of parameters and $\phi\simeq \pi$, this result yields
\begin{equation}
\label{supr}
|t(L)_{\downarrow}^{\downarrow}|^2\ll 1 \Longleftrightarrow N\gg \frac{1}{x\pi}\approx 45\,.
\end{equation}
The illustration of this phenomenon is displayed
in Fig.\ref{fig_RT200} for the transmission probabilities through 1, 20, 100 and 200
sequentially connected ripples ($\pi$-arcs).
Here, we consider the transmission as a function of the curved surface
energy $E_A$ of the incoming electrons (holes).
A small difference between spin up and spin down transmission probabilities
for one ripple (Fig.\ref{fig_RT200}a) at $E_A>0$ evolves to
$\simeq 100\%$ efficiency for  the spin up transmission probabilities
for the left-side incoming electron at $N=200$ ripples (Fig.\ref{fig_RT200}d).
The opposite picture takes place for
the spin down transmission probabilities at $E_A<0$.
To realize such a situation one might use the SiO$_2$ substrate as a gate
of the curved surface, which helps control the concentration of charge carriers in graphene.
As a result, one can change the charge carrier type from electron to hole \cite{Mor}.

\subsection{Spin filtering and  ripple parameters $R$ and $\phi$}

In light of the above analysis, without  loss of generality, we can consider
$m_s\phi<1$ in order to observe the suppression effect [see Eq.(\ref{supr})].
With the aid of Eq.(\ref{mv}), taking into account that
$\lambda_x\gg\lambda_y$, this requirement leads
to the following inequality
\begin{equation}
\label{rf}
\lambda_x-\frac{\gamma}{R\phi}<|E_A|<\lambda_x+\frac{\gamma}{R\phi}\,.
\end{equation}
To remain at the maximum, for example, the transmission probability $|t(L)_{\uparrow}^{\uparrow}|^2=1$, 
it is necessary to fulfill condition
(\ref{eff1}). As a result, in light of Eq.(\ref{mv})
and the condition $\lambda_x\gg\lambda_y$, taking into account Eq.(\ref{lxy}), one obtains
\begin{equation}
\label{r}
R\simeq \frac{\gamma}{|E_A|}(\frac{\pi}{\phi}-\beta)\,,\quad \beta=\frac{1+4\delta p}{2}\,.
\end{equation}
Combining this equation with Eq.(\ref{rf}), we have
\begin{equation}
\label{fi}
\frac{\pi-1}{2\beta}<\phi<\frac{\pi+1}{2\beta}\,.
\end{equation}
Thus, Eqs.(\ref{r}),(\ref{fi}) determine the region of feasibility of
the parameters $R$ and $\phi$, where the spin filtering effects could exist
at fixed system (graphene) parameters  such
as $\gamma$, $\delta$, and the electron energy $E_A$.
From this observation, two arguments  follow in favor
of our findings. First, even at $\phi\neq\phi_c$
(see Eq.(\ref{theta_c})) one of the  spin component in the
incoming electron (hole) flow is suppressed for a large enough
number of ripples at some particular  energy region.
Second, we assume that all ripples are identical.
Practically, the graphene surface is randomly curved, and it is
a real challenge to create identical, consequently connected ripples.
However, it is our belief that  modern technology will
allow us to realize this situation soon or later. Whatever the case,
the spin filtering effect should survive if small variations
of  radii and angles of consequently connected ripples  are 
subject to conditions (\ref{r}),(\ref{fi}), at a fixed  value $E_A$ of the electron energy flow.

\subsection{Effect of a finite $k_y$-momentum}

In our model a single ripple is modeled
as part of a nanotube that is infinite in the $y$ direction.
Evidently, realistic ripples are limited in space in both the $x-$ and $y-$ directions.
In particular, graphene nanoribbons are  considered  prominent candidates to control
the electronic properties of graphene based devices.
This issue requires, however, a dedicated study,
and is the subject of a forthcoming paper.

In order to have some idea of what should be expected
in graphene nanoribbons, we analyze the case with a finite
$k_y \neq 0$.
Nonzero $k_y$ could mimic the case of a ripple limited in the $y$-direction.
Indeed, a finite width in the $y$-direction introduces the quantization of
the $k_y$ momentum on the curved surface.
As a result, the eigenspinors at the curved
surface would depend on the mixture $\pm k_y$ values for $s=\pm 1$,
i.e., altogether four momentum $k_y$
(see details in Ref.\onlinecite{at1}). In this case
analytical expressions are too cumbersome,
even in  a simple case of one conserved  momentum $k_y$
on the curved surface.
Therefore, we proceed with a
numerical analysis that provides a vivid presentation of
a simple case with a single value of the $k_y$
momentum on the curved surface.

Let us suppose that the incoming electron flow
possesses a momentum $\vec{k}=(k_x,k_y)$  in regions L (R).
Evidently, in this case $E_0=\pm\gamma\sqrt{k_x^2+k_y^2}$.
For  simplicity, we consider $E_0>0$,
and obtain for the momentum on the curved surface
\begin{equation}
k_y=t_y/\gamma=E_0/\gamma\times \sin(\alpha)\,.
\end{equation}

The results of the calculations exhibit a degradation of the spin filter
ability of our system. At a fixed value of the energy $E_0=\lambda_x$
 the transmission probability $|t(L)_{\uparrow}^{\uparrow}|^2$
 decreases drastically at $|\alpha|\geq \pi/8$
[see Fig.\ref{fig_200ky}(a)]. It seems that the spin filtering
effects would survive at $|\alpha|< \pi/8$.
Note, however, that this estimation depends on the system
parameters, such as $\gamma$ and $E_0$.

At a fixed value of the momentum $k_y$
the  effectivity of  spin filtering is reduced by $\sim 10\%$
[see Fig.\ref{fig_200ky}(b)]. At the same time, our systems
manifests a zero transmission for all spin orientation
for charge carriers at energies $-0.06eV<E_A<0.06eV$ due to
our choice of the value $k_y$.

\begin{figure}
\includegraphics[height=5cm,clip=]{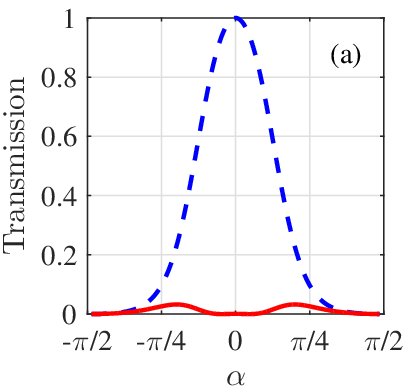}
\includegraphics[height=5cm,clip=]{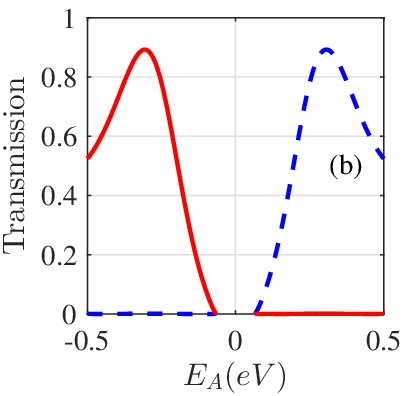}
\caption{(Color online) Transmission
probabilities $|t(L)_{\uparrow}^{\uparrow}|^2$ (blue, dash lines) and
$|t(L)_{\downarrow}^{\downarrow}|^2$ (red, solid lines):
(a) as a function of the incidence angle
$\alpha$ at $E_0=E_A=\lambda_x=0.32$eV;
(b)as a function of the energy $E_A$ at $k_y=0.01$\AA$^{-1}$.
The calculations are done for 200 sequentially connected ripples ($\pi$-arcs)
The other parameters are the same as in Fig.\protect\ref{fig_spectrum}.}
\label{fig_200ky}
\end{figure}

\subsection{The graphene  purity}

We restricted our consideration to a ballistic regime.
This approximation is well justified due to the following factors.
The remarkable strength of the carbon honeycomb lattice makes it
quite difficult to introduce any defects into the lattice itself.
Charge impurities that could limit electron mobility in graphene
are still  an open problem from both experimental and
theoretical points of view
(see, for example, discussion in Ref.\onlinecite{1}).
It is also well known that the difference in conductivity in graphene
between $T\approx 0$ and room temperature is no more than a few percent.
In other words, the electron-phonon scattering plays a minor role.

We recall that a typical
ripple size is $\sim 7$ nm (see Ref.\onlinecite{fas}). In our paper the ripple is modeled
as the curved surface in the form of an arc of a circle with a radius $R=1$nm.
As a result, our system length is $\pi R\times 200\approx 640$ nm. Taking
into account that a typical mean free path
of electrons in  single-wall nanotubes is $\ell\approx 1$ $\mu$m
(see, e.g., Ref.\onlinecite{2}), it seems our consideration is on a reasonable basis.

Thus, in our model the basic mechanism that is responsible for spin filtering effects is
an attenuation of one of the transmitting modes. It  transforms to the evanescent mode
in the energy gap created by the SOC in the curved
surface. The multiplicative action of a large enough
number of ripples suppresses this transmitting mode  at certain conditions
that provide a high efficiency  for the other one.

\section{Summary}
We have analysed the transmission and reflection of ballistic electron flow
through a ripple in an effective mass approximation, when only the interaction
between nearest neighbor atoms is taken into account.
In our consideration a ripple consists of
the curved surface in the form of an arc of a circle connected
from the left-hand and the right-hand sides to two semi-infinite flat graphene sheets.
Considering the curved surface as
a part of the armchair nanotube, we have shown that
the curvature induced spin-orbit coupling
yields a backscattering [see Eqs.(\ref{S3},\ref{S4})]
with spin inversion. This spin inversion is
caused by the spin-orbit term that breaks spin symmetry
(a spin projection on the symmetry axis) in the
effective Hamiltonian of the armchair CNT.

In the energy gap created by the curvature induced spin-orbit coupling
there is a preference
for  one spin orientation, depending on the direction of the electron flow
at  normal incidence.
The width of the energy gap depends
in inverse proportion on the radius of the ripple.
At this energy range the ripple
acts as a semipermeable membrane which is more transparent for the incoming electrons
with spin up from the left-hand side and  with spin down
from the right-hand side, and {\it vise versa} for the holes. In other words,
there is a precursor of {\it chiral} transmission of spin components of the incoming
electron (hole) flow at a fixed energy.
For one ripple system this effect is, however, small.
In order to enforce this effect, we extended our consideration to a
curved surface of the sinusoidal wave type with $N$ arcs.
This step is of crucial importance to suppress one of the spin
components and to support the  spin inversion symmetry for the transmission
probability. The larger the number of consistently connected ripples (arcs) is,
the stronger the dominance of a specific spin component is in comparison with
the other in the transmission from the same direction. There is a cooperative
effect of {\it chiral spin}  transmissions produced by a large number of ripples.
To trace the $N$-dependence
we have derived a formula for a composite
transmission probability for well-polarized spin components:
i) Eq.(\ref{SN1}) for spin up electrons;  ii) Eq.(\ref{SN2})
for spin down electrons.  Based on these results, we predict a strong
spin filtering effect for a sufficiently large number of
arcs in the rippled graphene system.
In contrast to the usual waveguide that guides optical or sound waves of
a chosen frequency in a well--defined direction,
our system guides spin electron (hole) waves with a well--defined polarization in
one or another direction at a certain energy. It seems, therefore, natural to name this
system  {\it chiral spinguide}.

We have considered only a curved surface that
owes its origin to an armchair nanotube.
Evidently, our model can be extended to other types of origins.
However, the corresponding analysis requires a separate study.
We also neglected the effective magnetic
field that arises from the dependence of the hopping parameter
$\gamma$ on the curvature (see discussion in \cite{kg}).
This effect influences the local density of
states \cite{Vozmediano}.  It can cause the localization of the
electrons on the boundary between flat graphene and the  curved surface, similar
to the boundary state for some types of carbon
nanoparticles \cite{Fujita}. As a result, it might affect the efficiency of
the {\it spinguide}.
 Last, but not least, many body effects such as electron-electron
interaction should be incorporated and analyzed as well.
It is especially noteworthy that
electron-electron interaction, designed in the form  of  a specific potential barrier
 on the graphene sheet \cite{wang}, leads to separation of spin-polarized states.
In fact, this result is in close agreement with our finding, obtained for one ripple.
As  mentioned above, the curvature induced SOC simulates a penetrable barrier preferable
for transmission of only one of two spin components, depending on the direction and energy 
of the incoming electron (hole) flow. It would be interesting to study the interplay 
between the SOC and electron-electron interaction on the electron transport in our system.
Evidently, this consideration would allow us to study in more detail the effect of
impurities on the electron mobility in our system.

In conclusion, the transparency and
the mathematical rigor of our results
provide  good grounds to believe that spin filtering effects found in
this paper, giving rise to a {\it chiral spinguide} phenomenon, will be observable in experiment.

\section*{Acknowledgments} M.P. and K.N.P. are grateful for the warm hospitality and creative atmosphere
at UIB and JINR.
This work was supported in part by RFBR Grant 14-02-00723
and Slovak Grant Agency VEGA Grant No. 2/0037/13.

\end{document}